\RequirePackage{fix-cm}
\documentclass[smallextended]{svjour3}       
\smartqed  
\usepackage{braket}
\usepackage{amsmath}
\usepackage{pgfplots}
\usepackage{csquotes}
\usepackage{float}
\usepackage{multirow}
\usepackage{dcolumn}
\usepackage[colorlinks=true, citecolor=blue]{hyperref}
\usepackage{hhline}
\usepackage{amssymb}
\begin{document}

\title{Experimental realization of quantum violation of entropic noncontextual inequality in four dimension using IBM quantum computer
}


\author{ Suvadeep Roy\and
        Bikash K. Behera \and Prasanta K. Panigrahi 
}


\institute{Suvadeep Roy\at
              Department of Physical Sciences, Indian Institute of Science Education and Research Kolkata, Mohanpur 741246, West Bengal, India\\
            \email{sr15ms116@iiserkol.ac.in}           
           \and
           Bikash K. Behera \at
               Department of Physical Sciences, Indian Institute of Science Education and Research Kolkata, Mohanpur 741246, West Bengal, India\\
            \email{bkb13ms061@iiserkol.ac.in}   
              \and 
              Prasanta K. Panigrahi \at   Department of Physical Sciences, Indian Institute of Science Education and Research Kolkata, Mohanpur 741246, West Bengal, India \\ \email{pprasanta@iiserkol.ac.in} 
}

\date{Received: date / Accepted: date}

\maketitle

\begin{abstract}
In quantum information theory, entropic inequalities act as the necessary and sufficient conditions to noncontextuality. Here, we first experimentally demonstrate the violation of the entropic noncontextual inequality in a four-level quantum system, by using the five-qubit IBM quantum computer. The experimental result disproves the existence of a local realist model. 
\end{abstract}


\section{Introduction}
 
\emph{Entropy} plays a key role in quantum information theory. This physical entity quantifies the amount of uncertainty or randomness in the state of a physical system. Due to randomness in the outcome, Einstein, Podolsky, and Rosen \cite{vnci_EPR1} arrived at the conclusion that nature can not be described completely by using $\psi$ function. However, the realist hidden variable models present a complete interpretation of the state function ($\psi$) of a system. Hence, each measured values of any dynamical variable can be predetermined by the corresponding hidden variables ($\lambda$). It is evident that, imposing constraints on the realist models leads to consistency with the experimental results of quantum mechanics (QM). Some examples of such constraints are the incompatibility between QM and the local realist models of quantum phenomena \cite{vnci_B1}, and the inconsistency between QM and the noncontextual realist (NCR) models (Bell-Kochen-Specker (BKS) theorem \cite{vnci_SPJ1}).

Let's consider the following observables $\hat{A}$, $\hat{B}$ and $\hat{C}$, where $\hat{A}$ commutes with $\hat{B}$ and $\hat{C}$ while $\hat{B}$ and $\hat{C}$ do not commute. The property that the measured statistics of $\hat{A}$ is independent of measurements of $\hat{B}$ or $\hat{C}$ concludes $\emph{noncontextuality}$ at the level of QM statistical results. Generally, NCR models support the extension of such context-independence, from the level of quantum statistical values to any individual measured value of a dynamical variable (predetermined by $\lambda$). In a given NCR model, let v(A), v(B) and v(C) be the individual measured values of $\hat{A}$, $\hat{B}$ and $\hat{C}$ respectively, as specified by a $\lambda$. In Refs. \cite{vnci_SPJ1,vnci_B3}, it is explicated that, NCR models assign invalid values for greater than two dimensional Hilbert space, even if considering all possible set of experiments. Thus, contextual hidden variable models are the necessary ingredients to reproduce the prediction of QM. 

A series of experiments based on quantum contextuality have been reported \cite{vnci_hasegawa3,vnci_nature,vnci_liu,vnci_ams}. An ingenious proof by Klyachko \emph{et al.} has been provided for a three level system \cite{vnci_KCS1} using only five observables, which later found experimental verification \cite{vnci_L1}. This scheme showed contextuality in three dimensions using the minimum number of observables. Entropic approach  for showing the quantum contextuality \cite{vnci_CF1,vnci_KRK1} has recently been used for testing local realism \cite{vnci_BC1}. This approach has also been utilized for illustrating the incompatibility \cite{vnci_CF1,vnci_KRK1} between the noncontextual realism and QM for three-level system. 

Here, we make use of a five-qubit real quantum computer developed by IBM (International Business Machines) corporation, using which a flurry of experiments \cite{vnci_IBM1,vnci_IBM2,vnci_IBM3,vnci_IBM4,vnci_IBM5,vnci_IBM6,vnci_IBM7,vnci_IBM8,vnci_IBM9,vnci_IBM10,vnci_IBM11,vnci_IBM12,vnci_IBM13,vnci_IBM14,vnci_IBM15,vnci_IBM16,vnci_IBM17,vnci_IBM18,vnci_IBM19,vnci_IBM20} have been realized. We prepare the necessary quantum circuit using the quantum computer to show the violation of entropic noncontextual inequality for a four-level system consisting of five observables. 

The Brief Report is organized as follows. Sec. \ref{vnci_II} provides a detailed theory and brief derivation of the \emph{noncontextual inequality}. Sec. \ref{vnci_III} explicates the necessary quantum circuits using the 5 qubit quantum processor named ibmqx4, to test the violation of entropic inequality and discusses the results. Following which, Sec. \ref{vnci_IV} concludes the paper by summarizing and providing future directions of our work.  

\section{Noncontextual Entropic Inequlity for Four-level System \label{vnci_II}}
  
Suppose, we would like to perform the measurement of the following observables $X_1,...X_{i},...X_{n}$ of a given system. Then this model can be considered to be noncontextual, if there exists a joint probability distribution $p(x_{1}, .x_{i}, .x_{n}|X_{1}, ...,X_{i}, ...X_{n})$, whose marginals can be expressed as, $p(x_i|X_i)=\sum_{i=1,..,i-1,i+1,n}p(x_1,...,x_i,...,x_n|X_1,...,X_i,...,X_n)$. The Shannon entropy for the above case can be defined as, 
\begin{eqnarray}
&&H(X_1, ..., X_{i}, . . . X_n)\\
\nonumber
&=&  -\Sigma_{x_1, ...,x_{i}, . .  x_n} {p(x_1, ..., x_{i}, . .  x_n) \ log_{2} p(x_1, ..., x_{i}, . . ., x_n)}
\end{eqnarray}

The two properties of Shannon entropy are given by \eqref{vnci_eq:1} and \eqref{vnci_eq:2}.
\begin{eqnarray}
\label{vnci_eq:1}
H(X_{i},X_{j})&=H(X_{i}|X_{j})+H(X_{j})&=H(X_{j}|X_{i})+H(X_{i})\nonumber\\
\end{eqnarray}
\begin{eqnarray}
\label{vnci_eq:2}
H(X_{i},X_{j})\leq H(X_{i})+H(X_{j})
\end{eqnarray} 
From \eqref{vnci_eq:1} and \eqref{vnci_eq:2}, we have
\begin{eqnarray}
\label{vnci_eq:3}
H(X_{i}|X_{j})\leq H(X_{i})
\end{eqnarray}
As $H(X_{j}|X_{i})\geq 0$ always holds, 
from \eqref{vnci_eq:1} and \eqref{vnci_eq:3}, we obtain
\begin{eqnarray}
\label{vnci_eq:5}
H(X_{i})\leq H(X_{i},X_{j})
\end{eqnarray}
For our experimental purpose, we have chosen a five observable system to realize in a five qubit quantum processor. Let's consider the following observables $X_{1},X_{2},X_{3},X_{4},X_{5}$, which are cyclically commuting. By using \eqref{vnci_eq:1}, the joint Shannon entropy for this system can be written as,
\begin{eqnarray}
\label{vnci_eq:6}
\nonumber
&&H(X_{1},X_{2},X_{3},X_{4},X_{5})\\
\nonumber
&&\hspace{8 mm}=H(X_{1}|X_{2},X_{3},X_{4},X_{5})+H(X_{2}|X_{3},X_{4},X_{5})\\
&&\hspace{12 mm}+H(X_{3}|X_{4},X_{5})+H(X_{4}|X_{5})+H(X_{5})
\end{eqnarray}

Hence, by using Eqs. \eqref{vnci_eq:1}, \eqref{vnci_eq:3}, \eqref{vnci_eq:5} and \eqref{vnci_eq:6}, the noncontextual entropic inequality is found to be \cite{vnci_SRM1}
\begin{eqnarray}
\label{vnci_eq:9}
M=H(X_{5}X_{1})-H(X_{1}X_{2})-H(X_{2}X_{3})-H(X_{3}X_{4})\nonumber\\-H(X_{4}X_{5})+H(X_{2})+H(X_{3})+H(X_{4}) \leq 0\nonumber\\ \end{eqnarray}
 
\section{Violation of Entropic noncontextual inequality at IBM QE for four level system \label{vnci_III}} 
\begin{figure*}
    \centering
    \includegraphics[width=\textwidth]{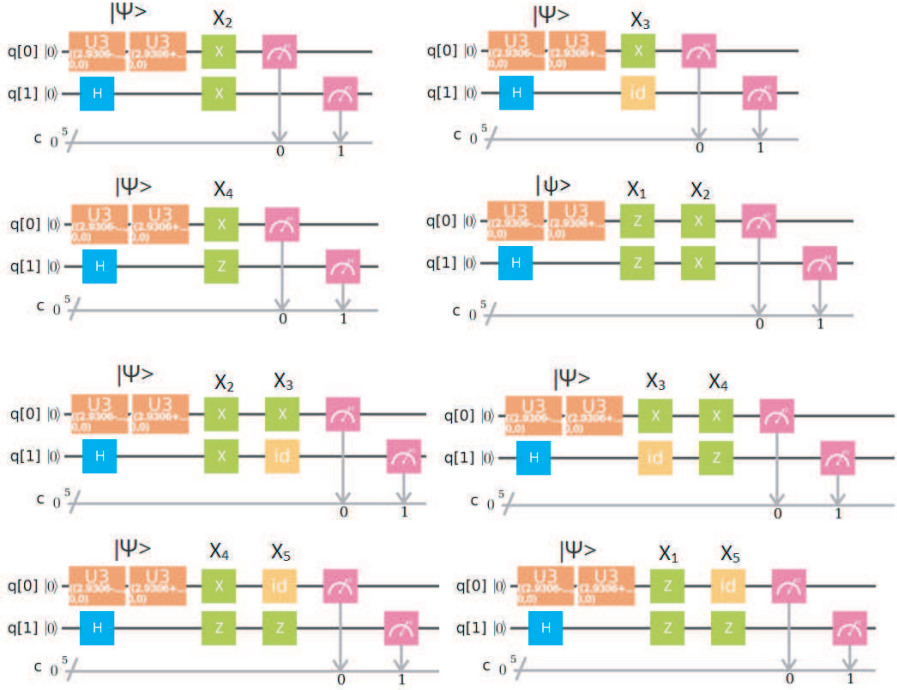}
    \caption{\emph{IBM quantum circuits used to measure Entropic noncontextual inequality by measuring seperately $H(X_{2})$,  $H(X_{3})$,  $H(X_{4})$,  $H(X_{1}X_{2})$,  $H(X_{2}X_{3})$,  $H(X_{3}X_{4})$,  $H(X_{4}X_{5})$ and  $H(X_{5}X_{1})$}for second set of observables as per the Table \ref{vnci_tabI}}
    \label{vnci_fig:I}
\end{figure*}
\begin{figure*}
    \centering
    \includegraphics[width=\textwidth]{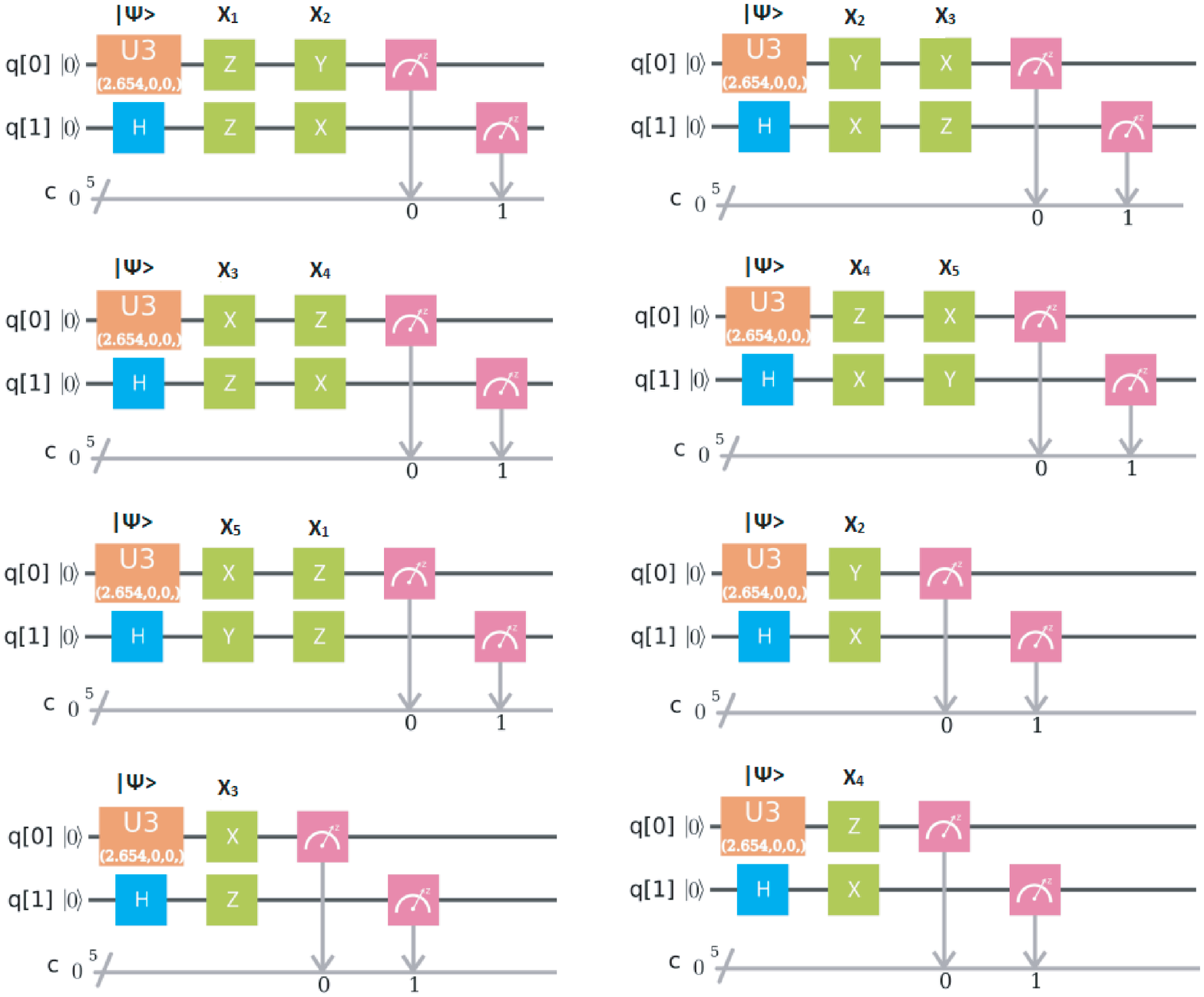}
    \caption{\emph{IBM quantum circuits used to measure Entropic noncontextual inequality by measuring seperately $H(X_{2})$,  $H(X_{3})$,  $H(X_{4})$,  $H(X_{1}X_{2})$,  $H(X_{2}X_{3})$,  $H(X_{3}X_{4})$,  $H(X_{4}X_{5})$ and  $H(X_{5}X_{1})$} for second set of observables as per the Table \ref{vnci_tabII}}
    \label{vnci_fig:II}
\end{figure*}
We have taken two sets of the cyclically commuting observables $X_{1},X_{2},X_{3},X_{4},X_{5}$ as per the Table \ref{vnci_tabI} and \ref{vnci_tabII} respectively.
\begin{table}[H] 
\centering
	\caption{\emph{The table illustrates one set of the chosen cyclically commutating observables.}}
	\begin{tabular}{cc}
		\hline
		\hline
		Observables
		&Form\\
		\hline
        $X_{1} $&$ \sigma_{Z}\otimes\sigma_{Z}$\\
		$X_{2} $&$ \sigma_{X}\otimes\sigma_{X}$\\
		$X_{3} $&$ \sigma_{X}\otimes I$\\
		$X_{4} $&$ \sigma_{X}\otimes\sigma_{Z}$\\
		$X_{5} $&$ I \otimes\sigma_{Z}$\\
		\hline
		\hline
	\end{tabular}
	\label{vnci_tabI}
\end{table}
\begin{table}[H] 
\centering
	\caption{\emph{The table illustrates the other set of the chosen cyclically commutating observables.}}
	\begin{tabular}{cc}
		\hline
		\hline
		Observables
		&Form\\
		\hline
        $X_{1} $&$ \sigma_{Z}\otimes\sigma_{Z}$\\
		$X_{2} $&$ \sigma_{Y}\otimes\sigma_{X}$\\
		$X_{3} $&$ \sigma_{X}\otimes\sigma_{Z}$\\
		$X_{4} $&$ \sigma_{Z}\otimes\sigma_{X}$\\
		$X_{5} $&$ \sigma_{X}\otimes\sigma_{Y}$\\
		\hline
		\hline
	\end{tabular}
	\label{vnci_tabII}
\end{table}

We have found the states $|\psi_{s}\rangle$= $C_{1}({cos\alpha,cos\alpha,sin\beta,sin\beta})^{T}$and $|\psi_{s}\rangle$=$C_{2}({sin\alpha,sin\alpha,cos\beta,cos\beta})^{T}$by using $U_{3}$ gates with specified values of $\lambda$, $\phi$ and $\theta$ and $C_{1},C_{2}$ as normalization constants.

The IBM quantum circuit for illustrating the violation of the non-contexual entropic inequality has been depicted in Fig. \ref{vnci_fig:I}. We have generated the sets of four level cyclically orthogonal observables as mentioned in Sec. \ref{vnci_II}. The experiment has been run with 8192 shots for the first set of observables with $\Ket{\psi_{s1}}$ for $\alpha$=$\beta$=2.9306 and for the other set of observables with $\Ket{\psi_{s2}}$ for $\alpha$=2.9306 and $\beta$=-5.7112. The corresponding results are obtained. The circuits are provided in the Figures \ref{vnci_fig:I} for $\ket{\psi_{s}}$ and the corresponding entropic calculations are shown in the Table \ref{vnci_tabIII} and \ref{vnci_tabIV}.
\begin{table}[H] 
\centering
\caption{\emph{The table depicts the entropic calculations on the experimental run (8192 shots) results for the $\Ket{\psi_{s1}}$ state with the first set of observables as shown in \ref{vnci_tabI}}}
\begin{tabular}{cc}
\hline
\hline
Entropy&Measured Value\\
\hline
$H(X_{2})$ & 1.64585197639\\
$H(X_{3})$ & 1.64895625081\\
$H(X_{4})$ & 1.59833444323\\
$H(X_{1}X_{2})$ & 1.66393718437\\
$H(X_{2}X_{3})$ & 1.27965313199\\
$H(X_{3}X_{4})$ & 1.28397144279\\
$H(X_{4}X_{5})$ & 1.63159920673\\
$H(X_{5}X_{1})$ & 1.28194875079\\
\hline
\hline
\label{vnci_tabIII}
\end{tabular}
\end{table}

\begin{table}[H] 
\centering
\caption{\emph{The table depicts the entropic calculations on the experimental run (8192 shots) results for the state $\Ket{\psi_{s2}}$ with the second set of observables as shown in \ref{vnci_tabII}.}}
\begin{tabular}{cc}
\hline
\hline
Entropy&Measured Value\\
\hline
$H(X_{2})$ & 1.06520690834\\
$H(X_{3})$ & 0.93645713795\\
$H(X_{4})$ & 1.13336434612\\
$H(X_{1}X_{2})$ & 0.96298009177\\
$H(X_{2}X_{3})$ & 1.09859136316\\
$H(X_{3}X_{4})$ & 0.93969773354\\
$H(X_{4}X_{5})$ & 0.96202918695\\
$H(X_{5}X_{1})$ & 0.95424133222\\
\hline
\hline
\label{vnci_tabIV}
\end{tabular}
\end{table}

The entropic noncontextual inequality in this experimental case is calculated and shown in Table \ref{vnci_tabIV}. However, it is state dependent as it violates only for some range of states. Similarly the state independent violation can be shown with nine observables. Here, we can conclude that noncontextual entropic inequality is succesfully violated. 

\begin{table}[H]
	\centering
	\caption{\emph{The table illustrates the final result of the entropic noncontextual inequality with the chosen observables and the chosen states.}}
	\begin{tabular}{cc}
		\hline
		\hline
		Entropic Noncontextual Inequality\\
		\hline
		$M_{|s1>}$ & 0.31094  \\
		$M_{|s2>}$ & 0.12597  \\
		\hline
		\hline
	\end{tabular}
	\label{vnci_tabV}
\end{table}

It is observed that the entropic noncontextual inequality for both the run result is highly violated. This certainly refutes the validity of hidden variable theory. 

\section{Conclusion \label{vnci_IV}}

To conclude, we have demonstrated here the violation of entropic noncontextual inequality with a two qubit quantum system consisting of five observables which acts as a four level system. We hope, our work can be extended to explicate the quantum violation of entropic noncontextual inequality for higher dimensional systems as well. It can be realized for nine observable systems for state independent violation. 
\section*{Acknowledgments}
\label{acknowledgments}
SR and BKB acknowledge the support of Inspire Fellowship awarded by DST, Govt. of India. The authors are extremely grateful to IBM team and IBM QE project. The discussions and opinions developed in this paper are only those of the authors and do not reflect the opinions of IBM or IBM Quantum Experience team.


\begin{thebibliography}{39}%
\makeatletter
\providecommand \@ifxundefined [1]{%
 \@ifx{#1\undefined}
}%
\providecommand \@ifnum [1]{%
 \ifnum #1\expandafter \@firstoftwo
 \else \expandafter \@secondoftwo
 \fi
 }%
\providecommand \@ifx [1]{%
 \ifx #1\expandafter \@firstoftwo
 \else \expandafter \@secondoftwo
 \fi
}%
\providecommand \natexlab [1]{#1}%
\providecommand \enquote  [1]{``#1''}%
\providecommand \bibnamefont  [1]{#1}%
\providecommand \bibfnamefont [1]{#1}%
\providecommand \citenamefont [1]{#1}%
\providecommand \href@noop [0]{\@secondoftwo}%
\providecommand \href [0]{\begingroup \@sanitize@url \@href}%
\providecommand \@href[1]{\@@startlink{#1}\@@href}%
\providecommand \@@href[1]{\endgroup#1\@@endlink}%
\providecommand \@sanitize@url [0]{\catcode `\\12\catcode `\$12\catcode
  `\&12\catcode `\#12\catcode `\^12\catcode `\_12\catcode `\%12\relax}%
\providecommand \@@startlink[1]{}%
\providecommand \@@endlink[0]{}%
\providecommand \url  [0]{\begingroup\@sanitize@url \@url }%
\providecommand \@url [1]{\endgroup\@href {#1}{\urlprefix }}%
\providecommand \urlprefix  [0]{URL }%
\providecommand \Eprint [0]{\href }%
\providecommand \doibase [0]{http://dx.doi.org/}%
\providecommand \selectlanguage [0]{\@gobble}%
\providecommand \bibinfo  [0]{\@secondoftwo}%
\providecommand \bibfield  [0]{\@secondoftwo}%
\providecommand \translation [1]{[#1]}%
\providecommand \BibitemOpen [0]{}%
\providecommand \bibitemStop [0]{}%
\providecommand \bibitemNoStop [0]{.\EOS\space}%
\providecommand \EOS [0]{\spacefactor3000\relax}%
\providecommand \BibitemShut  [1]{\csname bibitem#1\endcsname}%
\let\auto@bib@innerbib\@empty

\bibitem{vnci_EPR1}Einstein, A., Podolsky, B., Rosen, N.: Can quantum-mechanical description of physical reality be considered complete? Phys. Rev. \textbf{47}, 777 (1935) 
\bibitem{vnci_B1} Bell, J.~S.: On the Einstein Podolsky Rosen paradox. Physics \textbf{1}, 195 (1964)
\bibitem{vnci_SPJ1} Kochen, S., Specker, E.: The problem of hidden variables in quantum mechanics. J. Math. Mech. \textbf{17}, 59 (1967)
\bibitem{vnci_B3}Bell, J.~S.: On the problem of hidden variables in quantum mechanics. Rev. Mod. Phy. \textbf{38},447 (1966)
\bibitem{vnci_hasegawa3}Bartosik, H., Klepp, J., Schmitzer, C., Sponar, S., Cabello, A., Rauch, H., Hasegawa, Y.: Experimental test of quantum contextuality in neutron interferometry. Phys. Rev. Lett. \textbf{103}, 040403 (2009)
\bibitem{vnci_nature} Kirchmair, G. \emph{et al.}: State-independent experimental test of quantum contextuality. Nature \textbf{460}, 494 (2009)
\bibitem{vnci_liu} Liu, B.~H., Huang, Y.~F., Gong, Y.~X., Sun, F.~W., Zhang, Y.~S., Li, C.~F., Guo, G.~C.: Experimental demonstration of quantum contextuality with nonentangled photons. Phys. Rev. A \textbf{80}, 044101 (2009)
\bibitem{vnci_ams}Amselem, E., R\aa{}dmark, M., Bourennane, M., Cabello, A.: State-independent quantum contextuality with single photons. Phys. Rev. Lett. \textbf{103}, 160405 (2009)
\bibitem{vnci_KCS1} Klyachko, A.~A., Can, M.~A., Binicio\ifmmode \breve{g}\else \u{g}\fi{}lu, S., Shumovsky, A.~S.: Simple test for hidden variables in spin-1 systems. Phys. Rev. Lett. \textbf{101}, 020403 (2008)
\bibitem{vnci_L1} Lapkiewicz, R. \emph{et al.}: Experimental non-classicality of an indivisible quantum system. Nature \textbf{474}, 490 (2011)
\bibitem{vnci_CF1} Chaves, R., Fritz, T.: Entropic approach to local realism and noncontextuality. Phys. Rev. A \textbf{85}, 032113 (2012)
\bibitem{vnci_KRK1} Kurzynski, P., Ramanathan, R., Kaszlikowski, D.: Entropic test of quantum contextuality. Phys. Rev. Lett. \textbf{109}, 020404 (2012)
\bibitem{vnci_BC1} Braunstein, S.~L., Caves, C.~M.: Information-theoretic Bell inequalities. Phys. Rev. Lett. \textbf{61}, 662 (1988)
\bibitem{vnci_IBM1} Alsina, D., Latorre, J.~I.: Experimental test of Mermin inequalities on a five-qubit quantum computer. Phys. Rev. A \textbf{94}, 012314 (2016)
\bibitem{vnci_IBM2} Berta, M., Wehner, S., Wilde, M.~M.: Entropic uncertainty and measurement reversibility. New J. Phys. \textbf{18}, 073004 (2016)
\bibitem{vnci_IBM3} Wootton, J.~R.: Demonstrating non-Abelian braiding of surface code defects in a five qubit experiment. Quantum Sci. Technol. \textbf{2}, 015006 (2017)
\bibitem{vnci_IBM4} Behera, B.~K., Banerjee, A., Panigrahi, P.~K.: Experimental realization of quantum cheque using a five-qubit quantum computer. Quantum Inf. Process. \textbf{16}, 312 (2017)
\bibitem{vnci_IBM5} Kalra, A.~R., Prakash, S., Behera, B.~K., Panigrahi, P.~K.: Experimental demonstration of the no hiding theorem using a 5 qubit quantum computer \href{https://arxiv.org/abs/1707.09462}{arXiv:1707.09462 (2017)} 
\bibitem{vnci_IBM6} Ghosh, D., Agarwal, P., Pandey, P., Behera, B.~K., Panigrahi, P.~K.: Automated error correction in IBM quantum computer and explicit generalization. Quantum Inf. Process. \textbf{17}, 153 (2018)
\bibitem{vnci_IBM7} Gangopadhyay, S., Manabputra, Behera, B.~K., Panigrahi, P.~K.: Generalization and Demonstration of an Entanglement Based  Deutsch-Jozsa Like Algorithm Using a 5-Qubit Quantum Computer. Quantum Inf. Process. \textbf(17), 160 (2018) 
\bibitem{vnci_IBM8} Wootton, J.~R., Loss, D.: A repetition code of 15 qubits \href{https://arxiv.org/abs/1709.00990}{arXiv:1709.00990 (2017)}
\bibitem{vnci_IBM9} Huffman, E., Mizel, A.: Violation of noninvasive macrorealism by a superconducting qubit: Implementation of a Leggett-Garg test that addresses the clumsiness loophole. Phys. Rev. A \textbf{95} 032131 (2017)
\bibitem{vnci_IBM10} Sisodia, M., Shukla, A., Pathak, A.: Experimental realization of nondestructive discrimination of Bell states using a five-qubit quantum computer. Phys. Lett. A \textbf{9}, 50 (2016)
\bibitem{vnci_IBM11} Schuld, M., Fingerhuth, M., Petruccione, F.: Implementing a distance-based classifier with a quantum interference circuit. Euro. Phys. Lett. \textbf{119}, 6 (2017)
\bibitem{vnci_IBM12} Majumder, A., Mohapatra, S., Kumar, A.:Experimental realization of secure multiparty quantum summation using five-qubit IBM quantum computer on cloud. \href{https://arxiv.org/abs/1707.07460}{arXiv:1707.07460 (2017)}
\bibitem{vnci_IBM13} Kandala A., \emph{et al.}: Hardware-efficient variational quantum eigensolver for small molecules and quantum magnets. Nature \textbf{549}, 242 (2017)
\bibitem{vnci_IBM14}  Li, R., Alvarez-Rodriguez, U., Lamata, L., Solano, E. : Approximate quantum adders with genetic algorithms: An IBM quantum experience. Quantum Meas. Quantum Metrol. \textbf{4}, 1 (2017) 
\bibitem{vnci_IBM15} Sisodia, M., Shukla, A., Thapliyal, K., Pathak, A.: Design and experimental realization of an optimal scheme for teleportion of an n-qubit quantum state.\href{https://arxiv.org/abs/1704.05294}{arXiv:1704.05294 (2017)} 
\bibitem{vnci_IBM16} Vishnu, P.~K., Joy, D., Behera, B.~K., Panigrahi, P.~K.: Experimental demonstration of non-local controlled-unitary quantum gates using a five-qubit quantum computer \href{https://arxiv.org/abs/1709.05697}{arXiv:1709.05697 (2017)}
\bibitem{vnci_IBM17} Yalçınkaya, I., Gedik, Z.: Optimization and experimental realization of the quantum permutation algorithm. Phys. Rev. A \textbf{96}, 062339 (2017)
\bibitem{vnci_IBM18} Dash, A., Rout, S., Behera, B.~K., Panigrahi, P.~K.: Quantum locker Using a novel verification algorithm and its experimental realization in IBM quantum computer.\href{https://arxiv.org/abs/1710.05196}{arXiv:1710.05196 (2017)}
\bibitem{vnci_IBM19} Viyuela, O., Rivas, A., Gasparinetti, S., Wallraff, A., Filipp, S., Martin-Delgado, M.A.: Observation of topological Uhlmann phases with superconducting qubits. npj Quantum Inf. \textbf{4}, 10 (2018)
\bibitem{vnci_IBM20} Choo, K., Keyserlingk, C.~W.~v., Regnault, N., Neupert, T.: Measurement of the entanglement spectrum of a symmetry-protected topological state using the IBM quantum computer. \href{https://arxiv.org/abs/1804.09725}{arXiv:arXiv:1804.09725 (2018)}
\bibitem{vnci_SRM1} Pan, A.~K., Sumanth, M., Panigrahi, P.~K.: Quantum violation of entropic noncontextual inequality in four dimensions. Phys. Rev. A \textbf{87}, 014104 (2013)

\end{thebibliography}
\end{document}